\documentclass[hyper,letterpaper,12pt]{article}
\usepackage{a4wide}
\usepackage{amsmath}
\usepackage[
      colorlinks=true,
      linkcolor=blue,
      urlcolor=blue,
      filecolor=black,
      citecolor=blue,
            ]{hyperref}
\usepackage{color}
\usepackage{graphicx}
\usepackage{amsfonts}
\usepackage{amssymb}
\usepackage{epsfig}
\usepackage{subfigure}
\usepackage{amsmath}
\usepackage{latexsym}
\usepackage{mathrsfs}

\begin{document}
\title{\bf \Large Van der Waals-like phase transition from holographic entanglement entropy in Lorentz breaking massive gravity}

\author{\large~~
Xian-Ming Liu$^1$\footnote{E-mail: xianmingliu@mail.bnu.edu.cn}~,
Hong-Bo Shao$^{2}$\footnote{E-mail: priscillayao2@gmail.com}~,
Xiao-Xiong Zeng$^3$\footnote{E-mail: xxzeng@itp.ac.cn}~
~
\\
\\
\small $^1$ School of Science, Hubei University for Nationalities, Enshi, 445000, China\\
\small $^2$ College of Science, Agricultural University of Hebei, Baoding, 071000, China\\
\small $^3$ School of Material Science and Engineering, Chongqing Jiaotong University,\\
\small       Chongqing ~400074, China\\}
\maketitle

\begin{abstract}
\normalsize In this paper, phase transition of AdS black holes in lorentz breaking massive gravity has been studied in the framework of holography. We  find that there is a first order phase transition(FPT) and second order phase transition(SPT)  both in Bekenstein-Hawking entropy(BHE)-temperature plane and holographic  entanglement entropy(HEE)-temperature plane.  Furthermore, for the FPT, the equal area law is
checked and for the   SPT, the critical exponent of the heat capacity is also
computed. Our results confirm that the phase structure of  HEE
is similar to that of  BHE in lorentz breaking massive gravity, which  implies that  HEE and BHE have some potential underlying relationship.
\end{abstract}

\newpage


\section{Introduction}
\label{sec:intro}
The study of  HEE and quantum phase transitions of black holes have gained  a lot of  interest recent years. On one hand, HEE can be used as a perfect probe to study quantum information science \cite{Bennett1993,Bennett2000,Nielsen2000}, strongly correlated quantum systems \cite{Preskill2000,Osborne2001,Zanardi2002,1,2,3,4,5,Cai:2017ihd,Cai:2015cya} and Many-Body Systems \cite{Amico2008,Laflorencie2016}.
 On the other hand, investigation on  HEE of black holes may shed some light on understanding the nature of BHE\cite{Susskind1994,Solodukhin2011}.

Nearly ten years ago, a holographic derivation of the  HEE in conformal quantum field theories was proposed by Ryu and Takayanagi using the famous AdS/CFT correspondence \cite{Ryu-Takayanagi200601,Ryu-Takayanagi200602}. Recently the   HEE has been used as a probe to investigate the phase structure of the Reissner-Nordstrom AdS black hole\cite{Johnson2014}. The results showed that there is a Van der Waals-like(VDP) phase transition at the same critical temperature in both the fixed charge ensemble and chemical potential ensemble in the holographic  HEE-temperature plane. They also found that the   SPT occurring for the HEE at the same critical point as the  BHE with nearly the same critical exponent. This work was soon generalized to the extended phase space where the cosmological constant is considered as a thermodynamical variable\cite{Caceres2015}. Very recently, the equal area law of HEE was proved to hold for the FPT in the HEE-temperature plane \cite{Nguyen2015}. Based on \cite{Johnson2014},  VDP phase transition of   HEE in various AdS black holes have been studied in \cite{Zeng201601,Dey2016,Zeng201602,Zeng201603,Zeng201701,Mo201701,Song2017,Zeng:2017zlm,Li:2017gyc,Cadoni:2017fnd}. All of these works showed that the HEE undergoes the same  VDP phase transition as that of the  BHE.

 Massive gravity theories have been considerable interest recently. One of these reasons is that
   this alternative theories of gravity could explain the accelerated expansion of the universe without dark energy. The graviton behaves like a lattice excitation and exhibits a Drude peak in this theory. Current experimental data from the observation of gravitational waves by advanced LIGO require the graviton mass to be smaller than the inverse period of orbital motion of the binary system, that is $m_g<1.2\times10^{-22} eV/c^2$\cite{Abbott2016}. Another important reason for the interest in massive gravity is that the possibility of the mass  graviton could be help to understand the quantum  gravity effect. The first to introduce a mass to the graviton is in \cite{Fierz1939}. However this primitive linear  massive gravity theory contains the so called Boulware-Deser ghosts problem \cite{Boulware1972} that was solved by a nonlinear massive gravity theory \cite{Rham2010,Rham2011}, where the mass terms are obtained by introducing a reference metric. Recently Vegh proposed a new reference metric to describe a class of strongly interacting quantum field theories with broken translational symmetry in the holographic framework\cite{Vegh2013}. The recent progress in massive gravity can  be found in \cite{Rham2014,Hinterbichler2012}.

 Here, we consider AdS black holes in a class of Lorentz breaking massive gravity. In the massive gravity, the graviton acquires a mass by Lorentz symmetry breaking, which is very similar to the Higgs mechanism. A review of Lorentz violating massive gravity theory can be found in \cite{Dubovsky2004,Rubakov2008}.
 In this paper, we focus on the study of the  VDP phase transition of AdS black holes in Lorentz breaking massive gravity using the holographic  HEE. The main motivation of this paper is to explore whether the BHE phase transition can also be described by HEE in Lorentz breaking massive gravity. Firstly,  we would like to extend proposals in \cite{Johnson2014} to study  VDP phase transitions in AdS black hole with a spherical horizon in Lorentz-violating massive gravity with the HEE as a probe. What's more, we also would like to check the Maxwell's equal area law and critical exponent of the heat capacity, which are two characteristic quantities in  VDP phase transition.

The organization of this paper is as follows. In the next section, we shall provide a brief
review of the black hole solution in Lorentz breaking massive gravity firstly. Then  we will study the  VDP phase transitions and critical phenomena for the AdS black hole in the  BHE-temperature plane.  In Section 3, we mainly concentrate on the  VDP phase transition and critical phenomena in the framework of   HEE. The last section is devoted to our discussions and conclusions.

\section{ Phase transition and critical phenomena of AdS Black Holes in Lorentz breaking massive gravity}
\subsection{Review of  AdS Black Holes in Lorentz breaking massive gravity }
The four dimensional Lorentz breaking massive gravity can be obtained by adding non-derivative coupling scalar fields to the standard Einstein gravity theory. As a matter field is considered, the theory can be described by the following action \cite{Dubovsky2004,Rubakov2008}
\begin{equation}
  S=\int{d^4x\sqrt{-g}[-M^2_{Pl}R+L_{m}+\ell^4 \Psi(X,\Pi^{ij})]},
\end{equation}
here the first two terms are  the curvature and ordinary matter minimally coupled to gravity respectively, the third term $\Psi$ contains two functions $X$ and $\Pi^{ij}$ which relate to the four scalar fields, $\Xi^0$ and $\Xi^i$ as
\begin{eqnarray}
  X&=&\frac{\partial^{\mu}\Xi^0\partial_\mu \Xi^0}{\ell^4},\\
  \Pi^{ij}&=&\frac{\partial^{\mu}\Xi^i\partial_\mu \Xi^j}{\ell^4}-\frac{\partial^{\mu}\Xi^i\partial_\mu \Xi^0\partial^{\nu}\Xi^j\partial_\nu \Xi^0}{\ell^8 X}.
\end{eqnarray}
When the four scalar fields  get a space-time depending vacuum expectation value, the system will break the Lorentz symmetry. What's more, the action  can also be taken as a low-energy effective theory with the ultraviolet cutoff scale $\ell$. Here the scale parameter $\ell$ has the dimensions of mass and is in the order of $\sqrt{m_g M_{Pl}}$, where $m_g$ and $M_{Pl}$ are the graviton mass and the Plank mass respectively.

\begin{figure}[tbp]
 \centering
\includegraphics[width=0.5\textwidth]{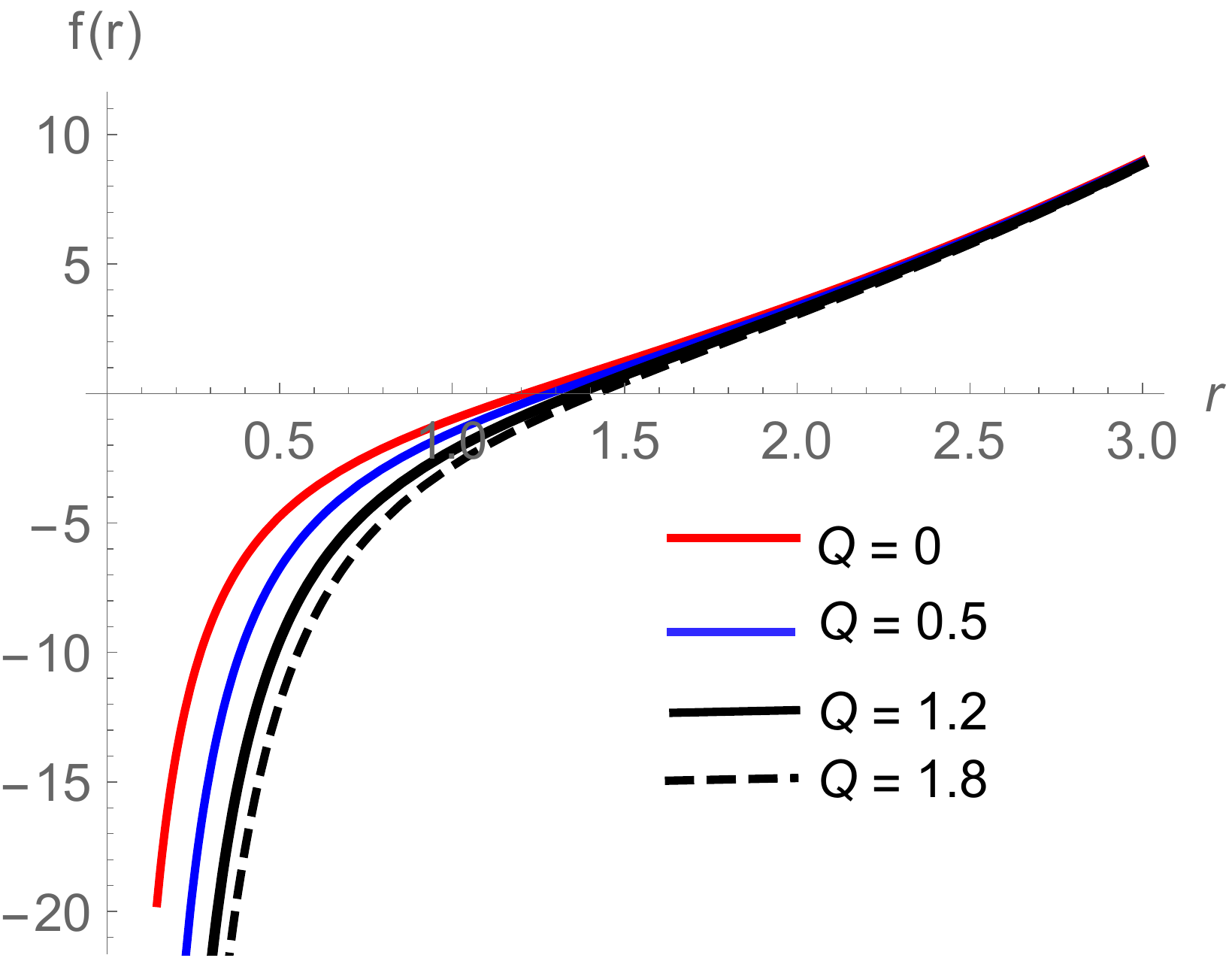}
 \caption{The figure shows f(r) vs r for $\gamma =1$ for varying Q. Here $\lambda =2.4$, $\Lambda =-3$ and $ M=1.5$.}
\label{fig:L1}
\end{figure}

\begin{figure}
\centering
\includegraphics[width=0.5\textwidth]{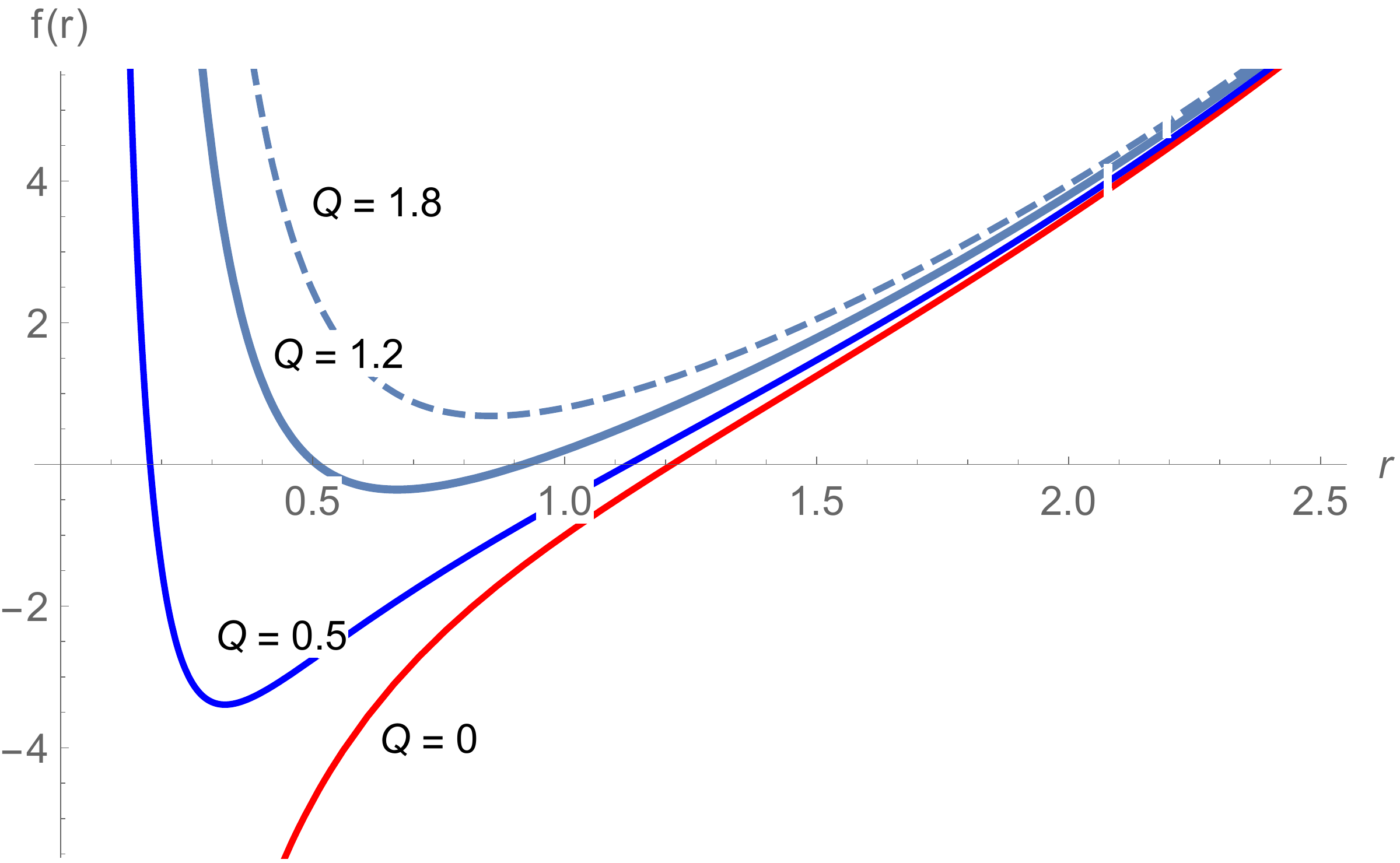}
 \caption{The figure shows f(r) vs r for $\gamma =-1$ for varying Q. Here $\lambda =2.4$, $\Lambda =-3$ and $ M=1.5$.}
\label{fig:L2}
\end{figure}

The AdS black holes solution can be obtained from the above theory\cite{Bebronne2009,Comelli2011}. The metric corresponding to the AdS black holes is given by
\begin{eqnarray}
  ds^2&=&-f(r)dt^2+f(r)^{-1}dr^2+r^2 (d\varphi^2+\sin^2\varphi d\phi^2),\label{2.01}
  \end{eqnarray}
  with
  \begin{equation}
    f(r)=1-\frac{2M}{r}-\gamma\frac{Q^2}{ r^\lambda}-\frac{\Lambda r^2}{3}. \label{2.02}
  \end{equation}
Here, the four scalar fields, $\Xi^0$ and $\Xi^i$, for this particular solution are given by
\begin{eqnarray}
  \Xi^0&=&\ell^2(t+\eta(r)),\\
  \Xi^i&=&\ell^2 \alpha x^i,
\end{eqnarray}
in which
\begin{equation}
  \eta(r)=\pm\int\frac{dr}{f(r)}[1-f(r)(\frac{\gamma Q^2 \lambda(\lambda-1)}{12m_g^2\alpha^6}\frac{1}{r^{\lambda+1}}+1)^{-1}]^{1/2},
\end{equation}
in which  the scalar charge Q is related to massive gravity and the constant $\alpha$ which  is determined by the cosmological constant $\Lambda$ and the graviton mass $m_g$  with the relation $\Lambda=2m_g^2(1-\alpha^6)$. In this paper we will set $\alpha>1 $ such that $\Lambda<0$ leading to Anti-de Sitter black holes.
The constant $\lambda$ is a positive integration constant. When $\lambda<1$, the
ADM mass of the black hole solution diverges. For $\lambda>1$, the metric approaches the Schwarzschild-AdS black holes with a finite mass M as $r\rightarrow\infty$. Thus we set $\lambda>1$ in this paper. The constant $\gamma=\pm 1$. When $\gamma=1$, the  black hole only has a single horizon $r_h$, which is the root of the equation $f(r_h) = 0$. The function $f(r)$ for this case is given in Fig.(\ref{fig:L1}), which is similar to the Schwarzschild AdS black hole. For $\gamma=-1$, the black hole is very similar to the Reissner-Nordstrom-AdS black hole. The function $f(r)$ for this case is given in Fig.(\ref{fig:L2}). The black hole event horizon $r_h$ is the largest root of the equation $f(r_h) = 0$.

 \begin{figure}[tbp]
 \centering
\includegraphics[width=0.5\textwidth]{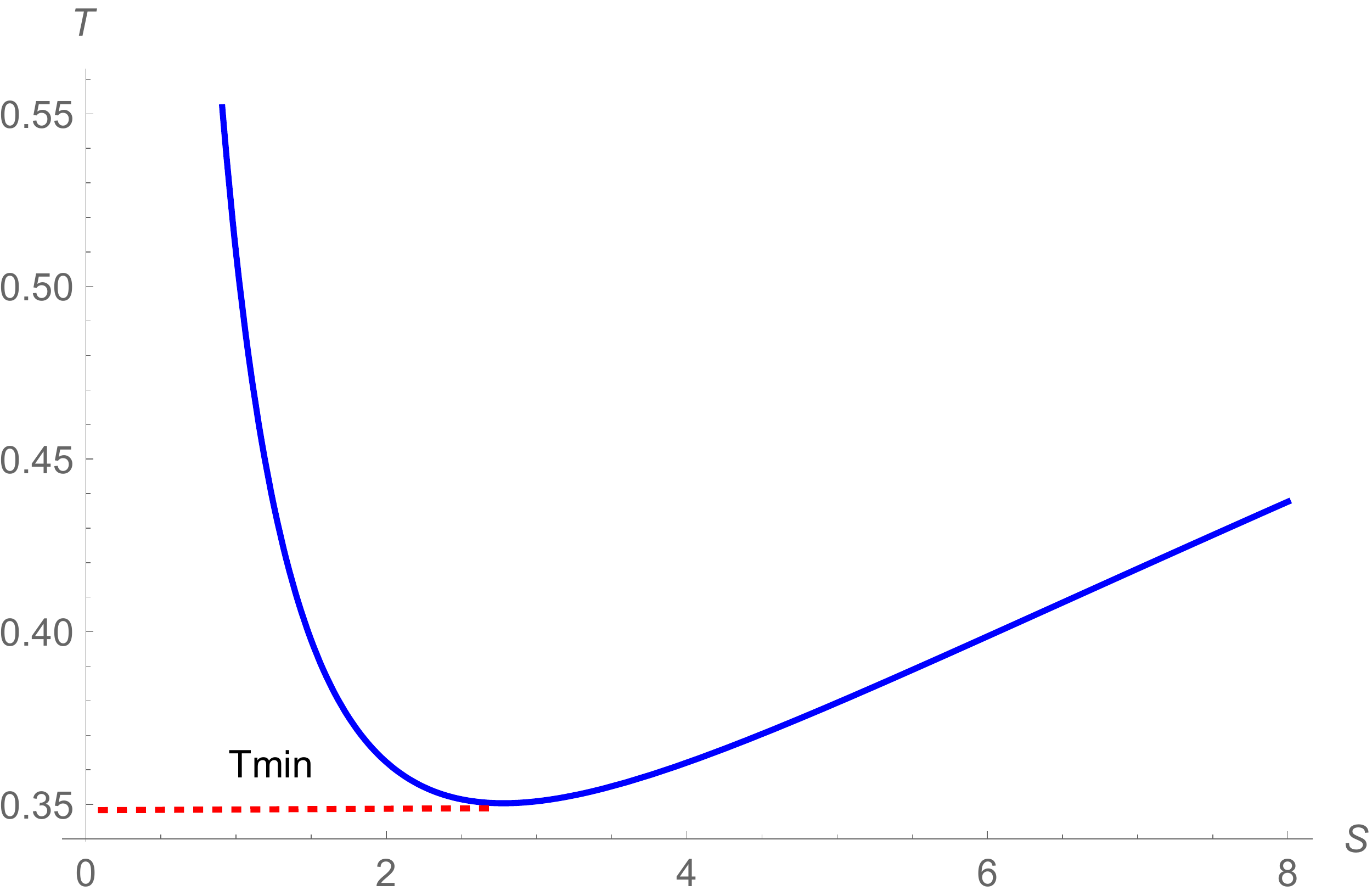}
 \caption{The figure shows T vs S for $\gamma =1$. Here $\lambda =2.4$, $\Lambda =-3$ and $ Q=0.3$.}
\label{fig:L3}
\end{figure}

 \begin{figure}[tbp]
 \centering
\includegraphics[width=0.5\textwidth]{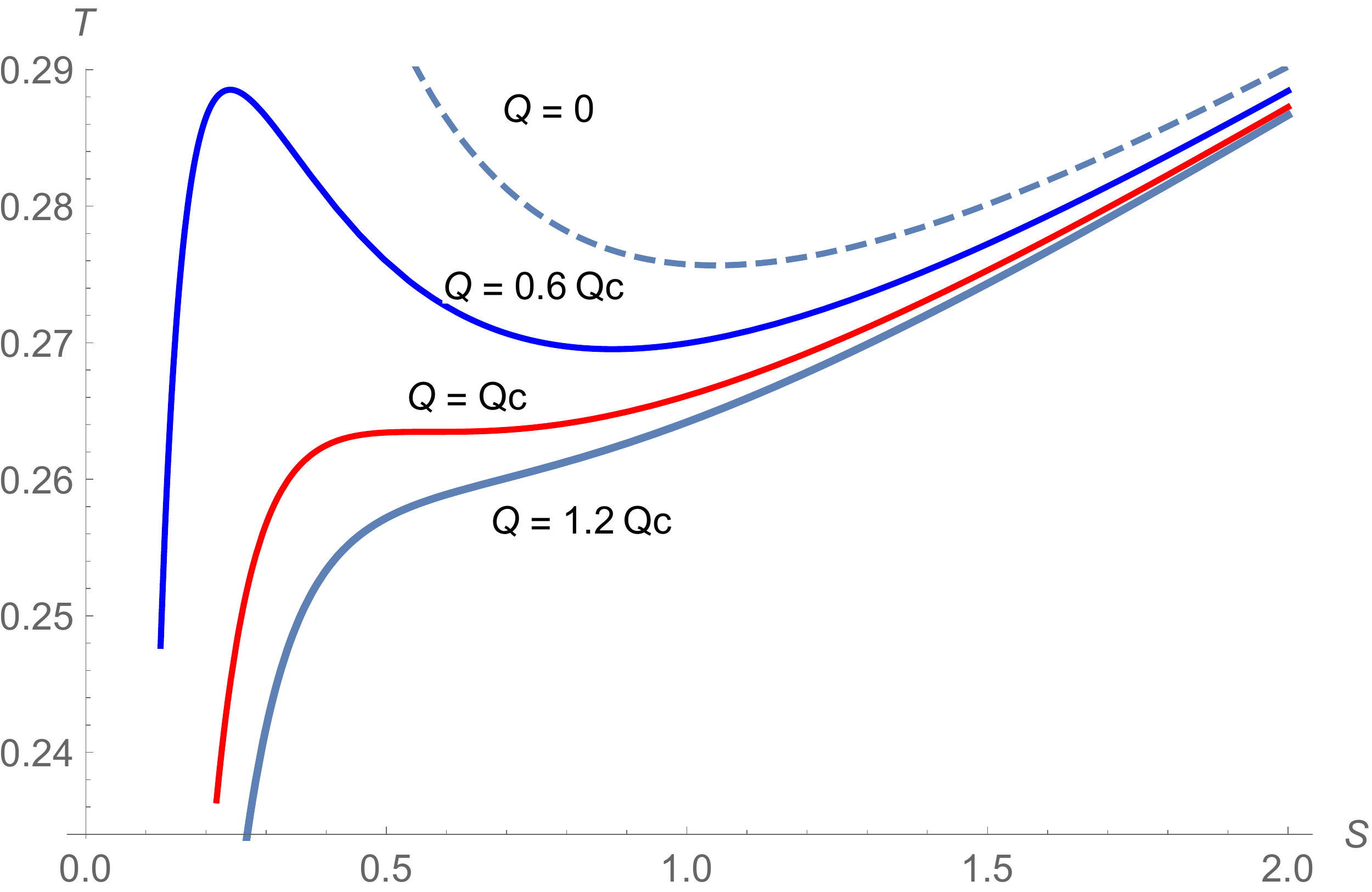}
 \caption{The figure shows T vs S for $\gamma =-1$ for varing Q. Here $\lambda =2.4$, $\Lambda =-3$. The top  dashed curve is at $Q=0$ and the rest have $Q=0.6Q_c, Q_c,1,2Q_c$}
\label{fig:L4}
\end{figure}
\begin{figure}[tbp]
\centering
\includegraphics[width=.3\textwidth]{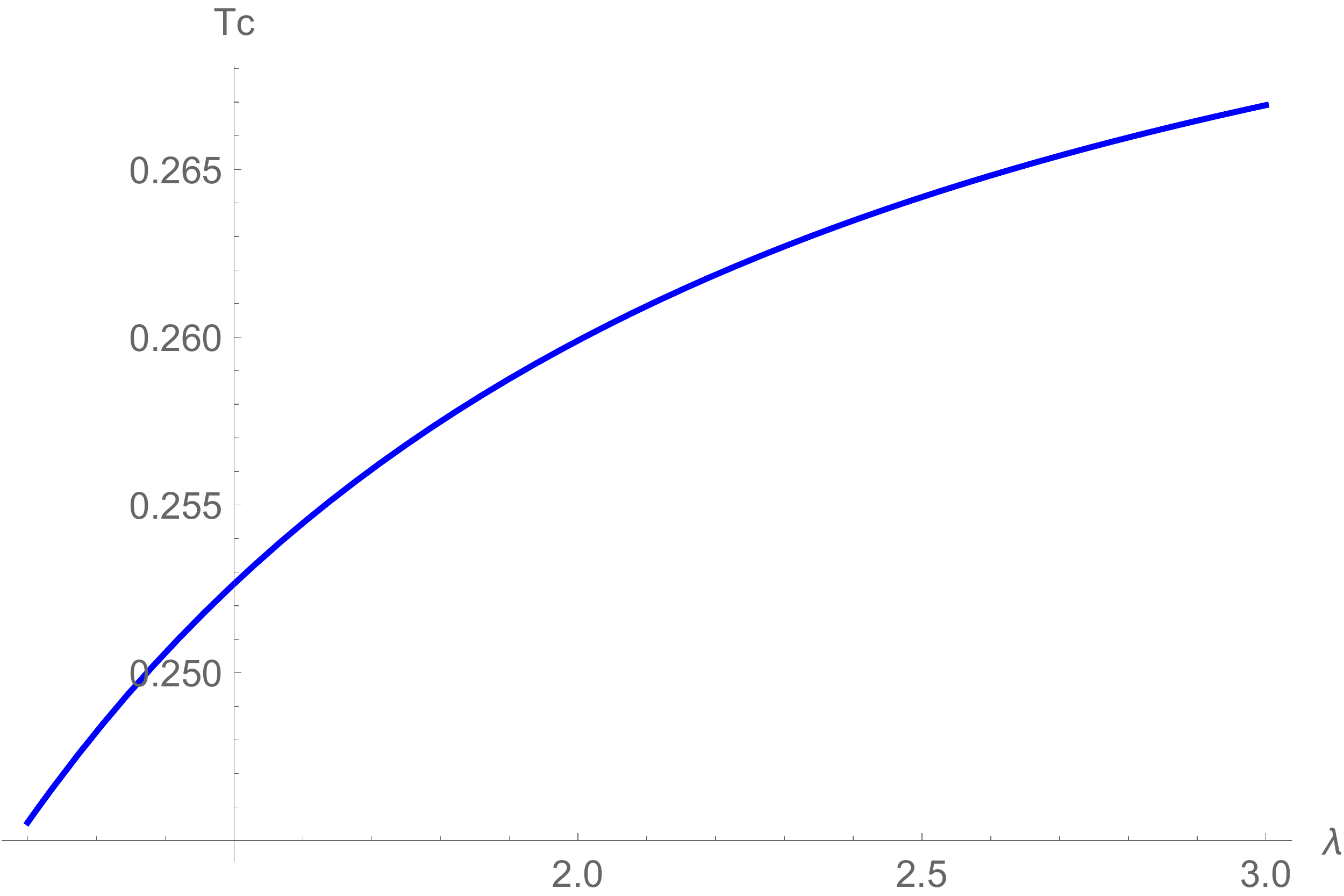}
\hfill
\includegraphics[width=.3\textwidth]{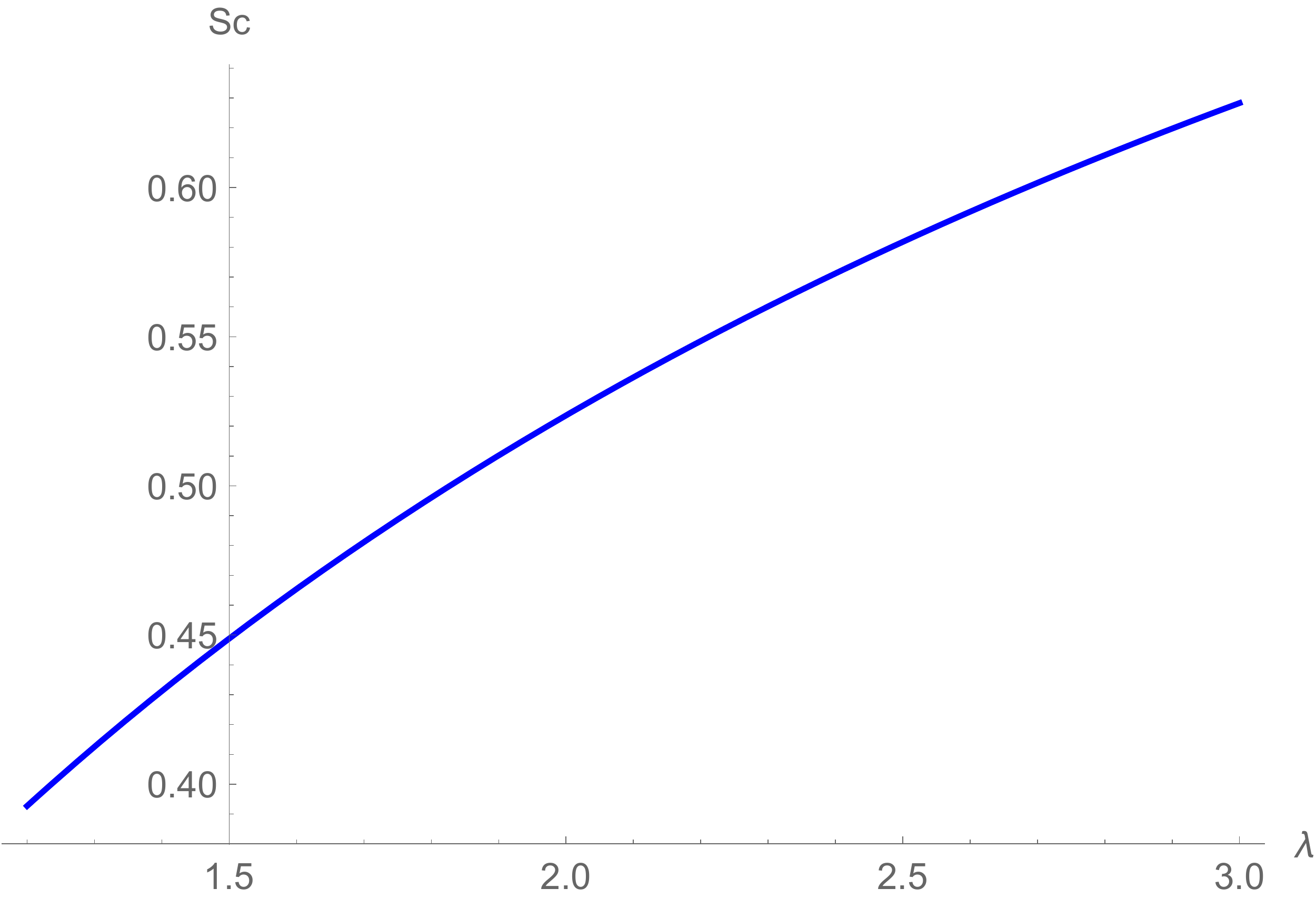}
\hfill
\includegraphics[width=.3\textwidth]{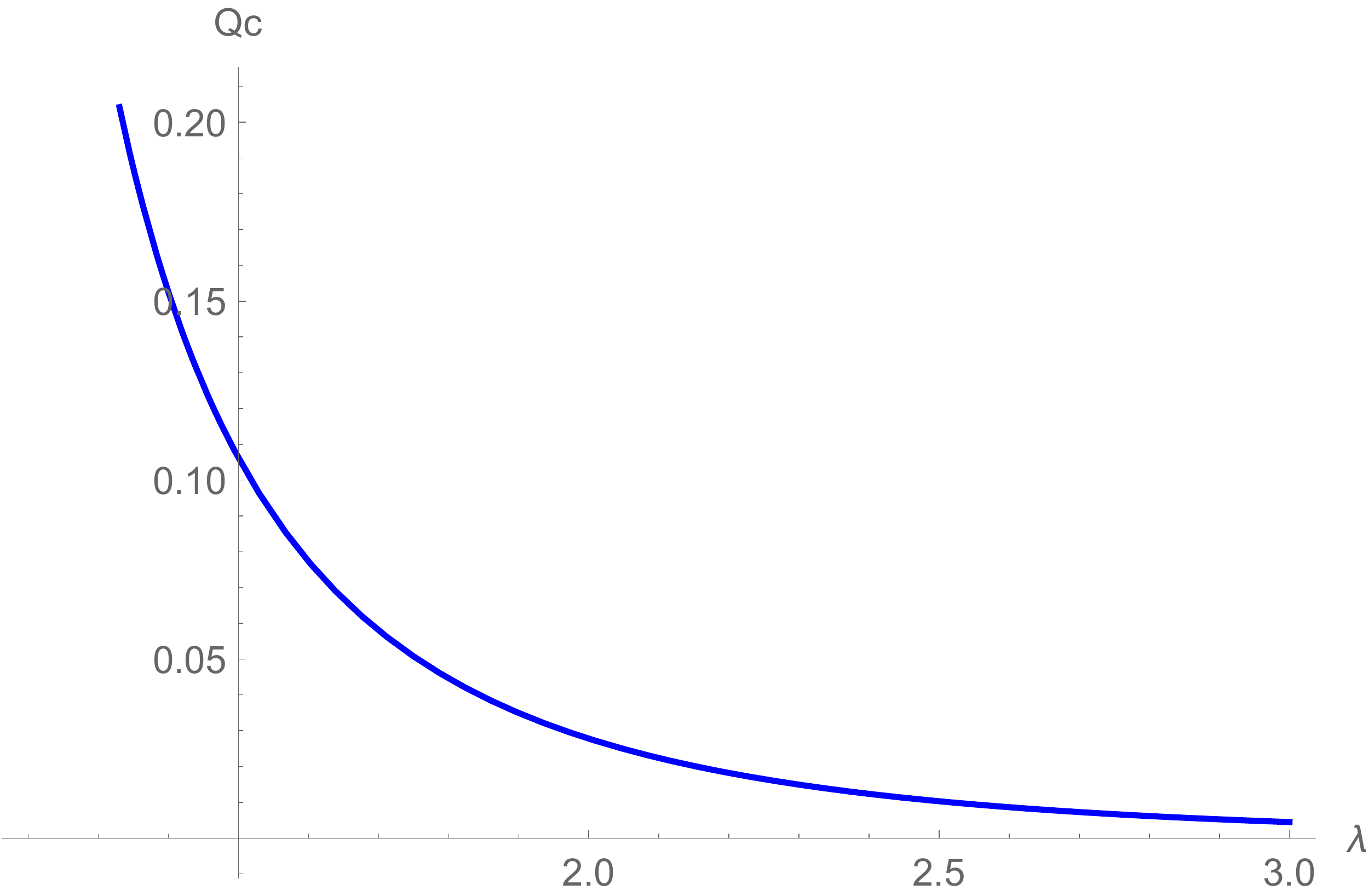}
 \caption{The figure shows the critical temperature $T_c$ , the critical entropy $S_c$ and the critical charge $Q_c$ vs  $\lambda$ for $\gamma =-1$ . Here $\Lambda =-3$.}
\label{fig:L7}
\end{figure}
\begin{figure}[tbp]
\centering
\includegraphics[width=.3\textwidth]{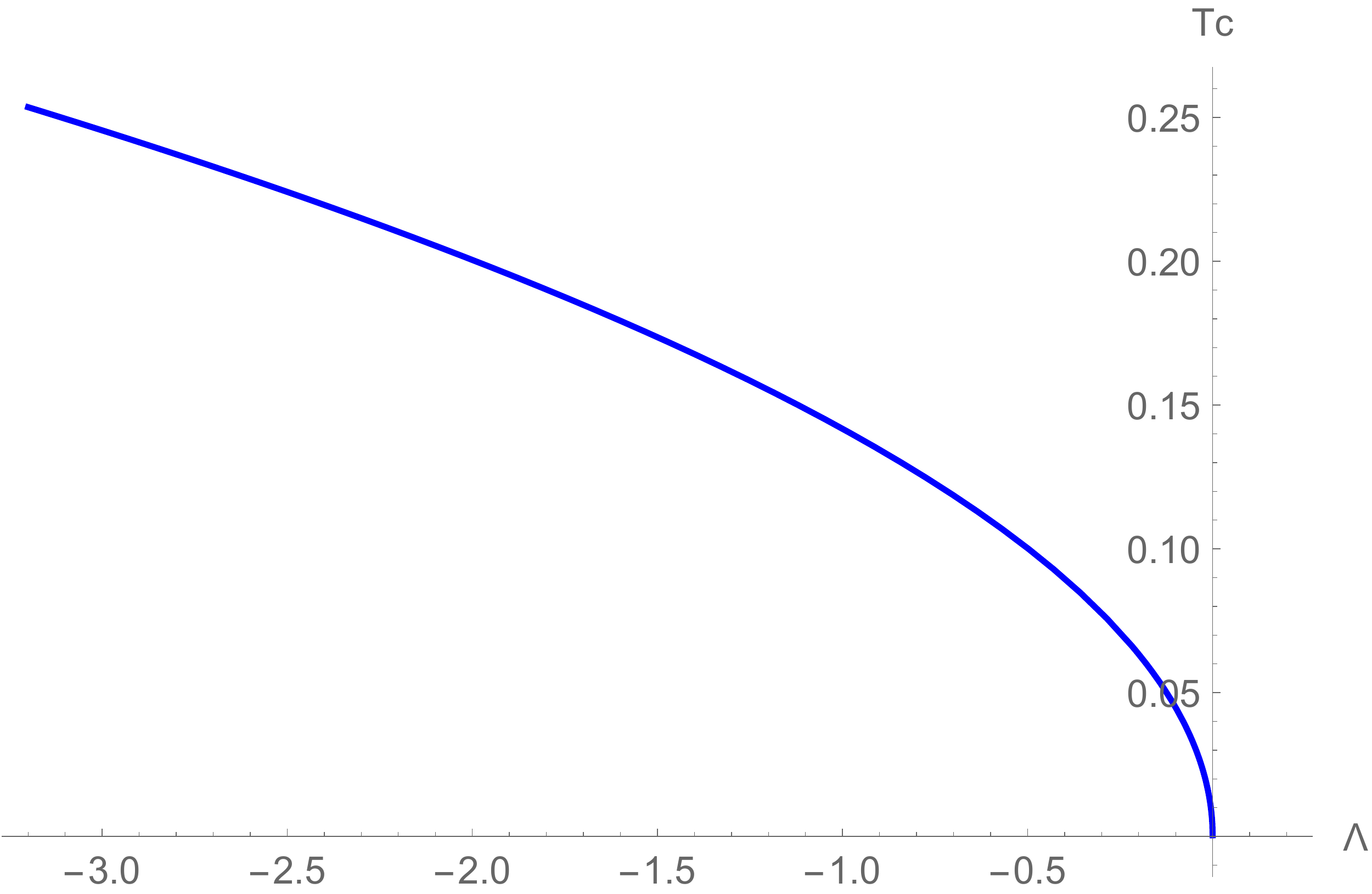}
\hfill
\includegraphics[width=.3\textwidth]{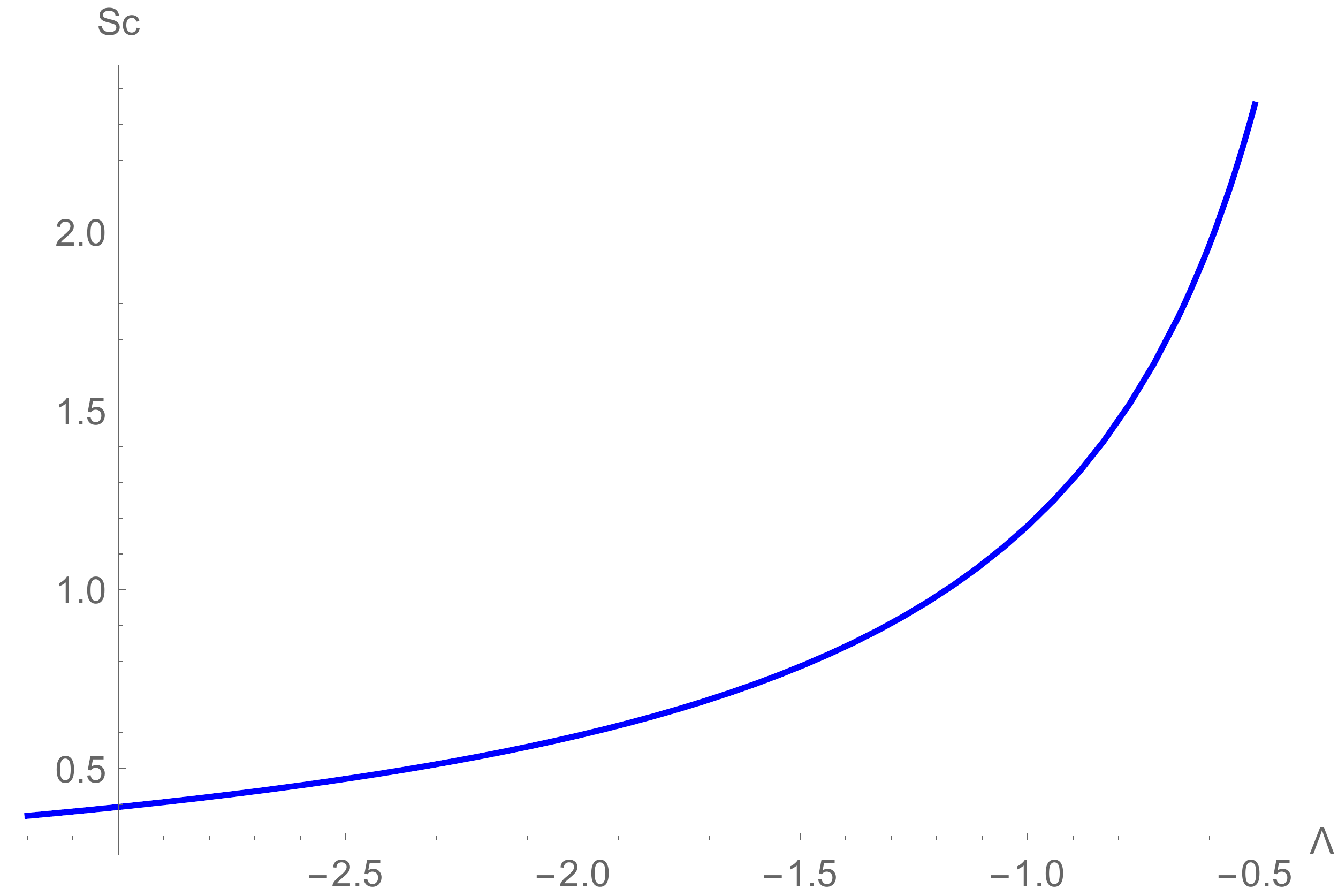}
\hfill
\includegraphics[width=.3\textwidth]{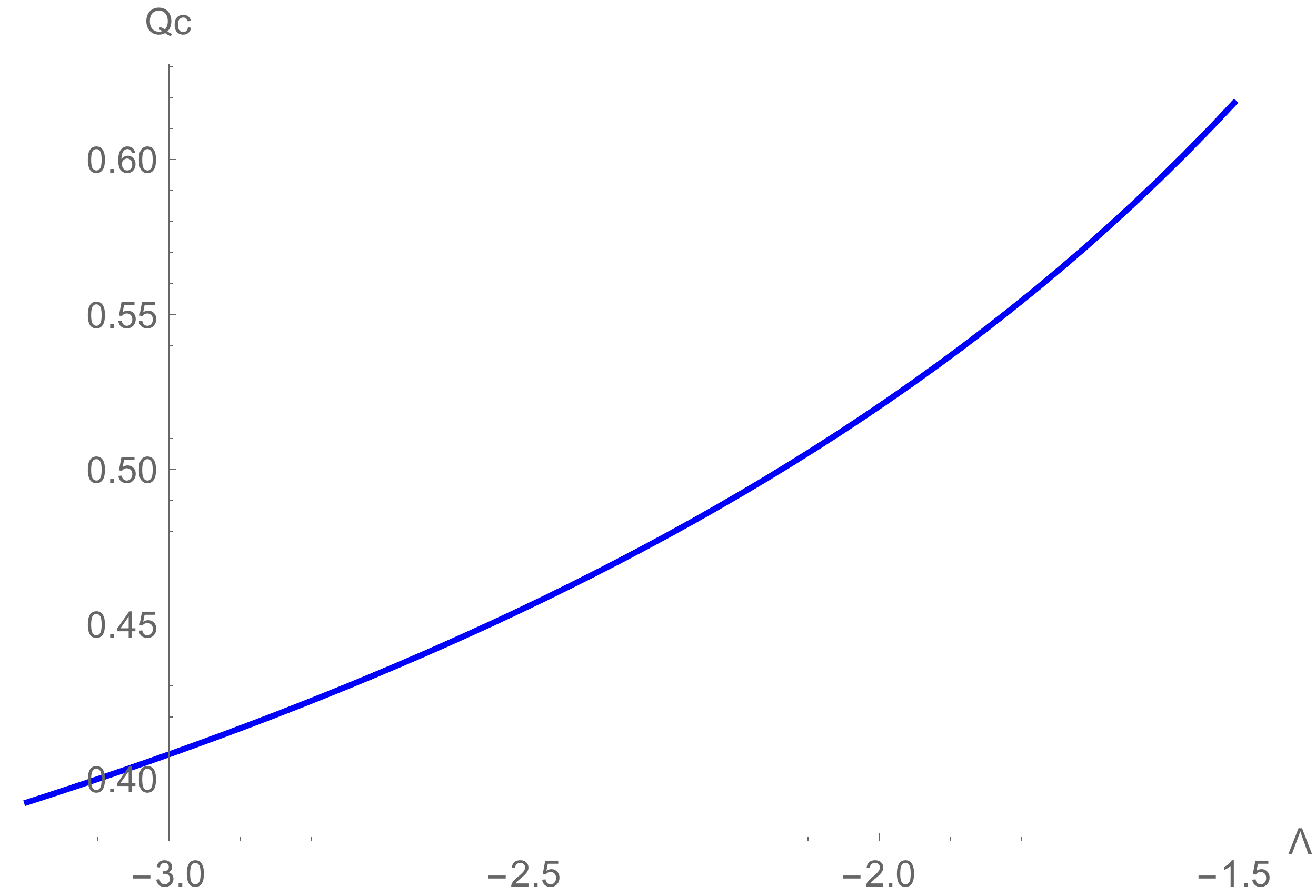}
 \caption{The figure shows the critical temperature $T_c$ , the critical entropy $S_c$ and the critical charge $Q_c$ vs $\Lambda$ for $\gamma =-1$. Here $\lambda =1.2$.}
\label{fig:L8}
\end{figure}
 \begin{figure}[tbp]
 \centering
\includegraphics[width=0.5\textwidth]{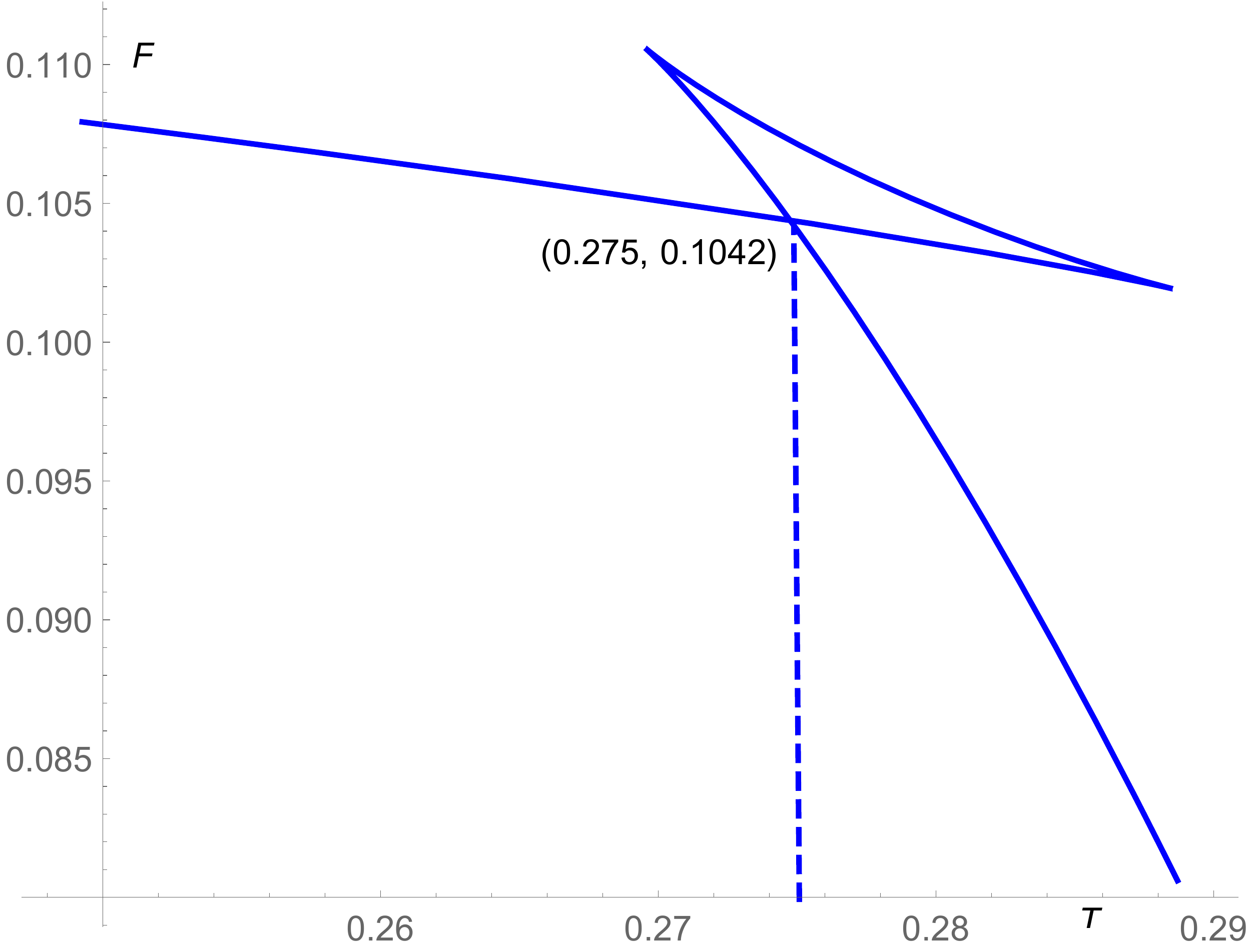}
 \caption{The figure shows F vs T for $\gamma =-1$. Here $\lambda =2.4$, $\Lambda =-3$,$Q=0.6 Q_c$.}
\label{fig:L5}
\end{figure}
 \begin{figure}[tbp]
 \centering
\includegraphics[width=0.5\textwidth]{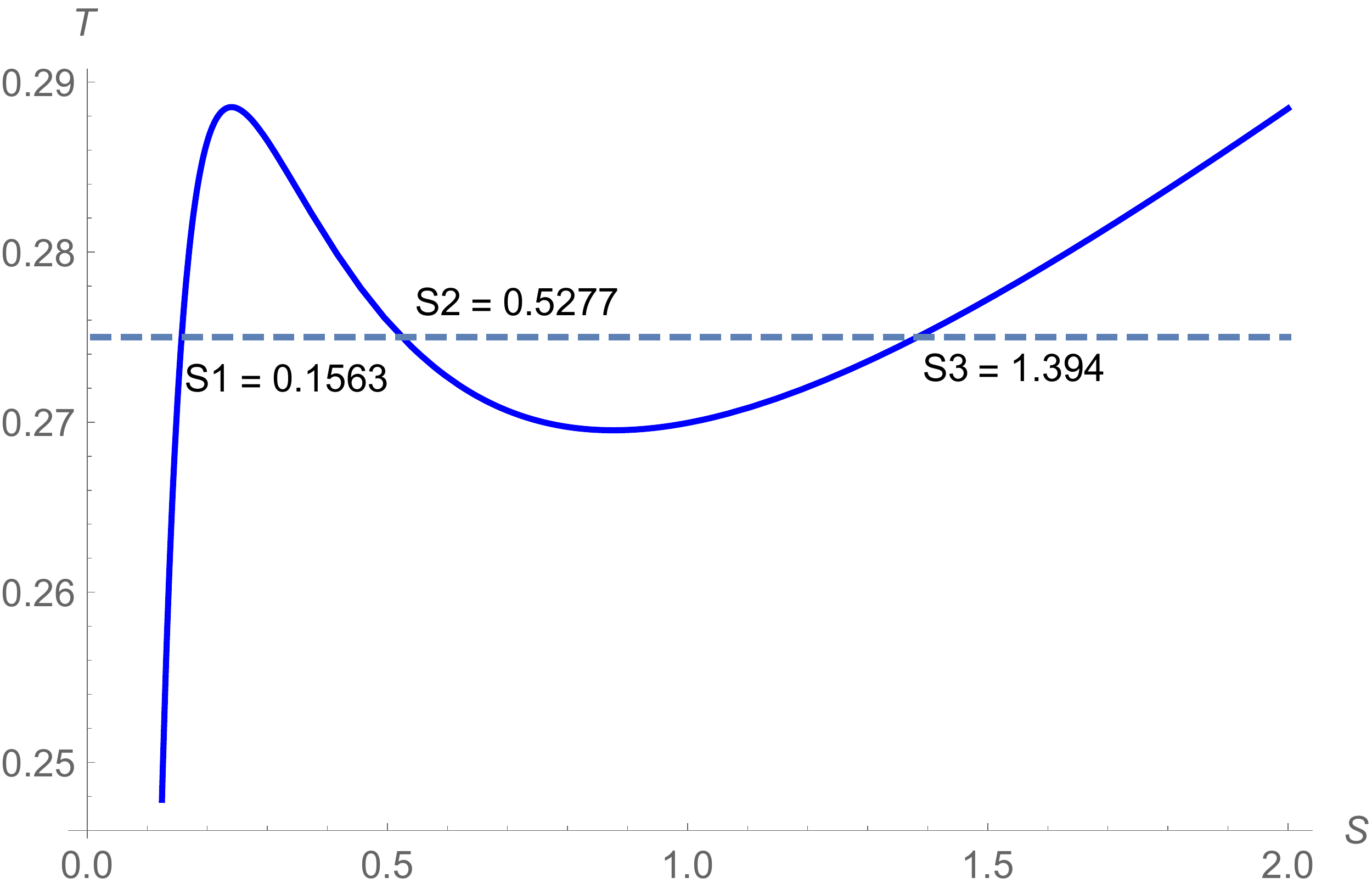}
 \caption{The figure shows T vs S for $\gamma =-1$. Here $\lambda =2.4$, $\Lambda =-3$,$Q=0.6 Q_c$.}
\label{fig:L6}
\end{figure}
At the event horizon, the Hawking temperature and   BHE can be written as
 \begin{equation}
 \label{2.9}
    T=\frac{1}{4\pi}f'(r_{h})=\frac{1}{4\pi}(\frac{1}{r_h}-r_h\Lambda+\frac{\gamma(\lambda-1)Q^2}{r_h^{\lambda+1}}),
 \end{equation}
  \begin{equation}
  \label{2.10}
   S=\pi r^2_{h}.
  \end{equation}
The chemical potential in this black hole is
\begin{equation}
  \Phi=-\gamma\frac{Q}{r_h^{\lambda-1}}.
\end{equation}
We can check the first law of the black hole, which is given by
\begin{equation}
  dM=T dS+\Phi dQ.
\end{equation}
There have been some works to study the thermodynamics and phase transitions of black holes in Lorentz breaking massive gravity \cite{Fernando2016,Capela2012,Mirza2014,Fernando201602}.

\subsection{Van der Waals-like Phase transition of  Bekenstein Hawking entropy}
In this subsection, we focus on the  VDP phase transition of  BHE.
Substituting \eqref{2.10} into \eqref{2.9} and eliminating the parameter $r_h$, one can get the relation between the Hawking temperature $T$ and  BHE $S$ of the AdS black holes in massive gravity as
\begin{equation}
\label{2.11}
 T=\frac{S^{\frac{1}{2} (-\lambda -1)} \left(\gamma  \pi ^{\frac{\lambda }{2}+1} (\lambda -1) Q-\Lambda  S^{\frac{\lambda }{2}+1}+\pi  S^{\lambda /2}\right)}{4 \pi ^{3/2}},
\end{equation}
This is the state equation of the AdS black hole thermodynamics system in massive gravity. Using \eqref{2.11}, we investigate the phase diagram of the AdS black holes in massive gravity.

The temperature $T$ is plotted as a function of the  BHE $S$ in Fig.(\ref{fig:L3})  and Fig.(\ref{fig:L4})  for $\gamma=1$ and $\gamma=-1$ respectively. In Fig.(\ref{fig:L3}), the temperature is plotted for $\gamma=1 $ where only one event horizon exists. This behavior of temperature is very similar to the behavior in the Schwazschild-AdS black hole. That is to say there is a minimum temperature $T_{min}$  which divides the thermodynamics systems into small and large black holes. It is shown that above the minimum temperature, small and large black holes coexist. In fact, this behavior will break for there is a first order transition, which is similar to the Hawking-Page thermodynamic transition in Ref.\cite{Hawking1983}.

In Fig.(\ref{fig:L4}), the temperature is plotted for $\gamma=-1$ where event horizon and inner horizon
exist. Various values of $Q$ is used to plot the relations between the temperature and horizons. The
top curve corresponds to $Q = 0$. The system of this case is similar to the case $\gamma= 1$  described above. When the scalar charge $Q$ increases, the temperature has two turning points. Further increasing of the scalar charge $Q$ brings these two turning points merge to one. It is shown that there exist a critical point $Q_c$. Above the critical point, the curve does not have any turning points. Thus we find that the phase structure is very similar to that of the Van der Waals gas-fluid phase transition. It should be noted that we  mainly concentrate on this type of phase structure in this paper. Furthermore using
the definition of the specific heat capacity $ C_Q$, that is
\begin{equation}
\label{2.14}
  C_Q=T(\frac{\partial S}{\partial T})_Q,
\end{equation}
one can see that the specific heat capacity is divergent at the critical point and it is obvious that this phase transition is a   SPT. At this critical point, the critical charge $Q_c$ and critical entropy $S_c$ can be obtained by the following equations
\begin{equation}
\label{2.15}
  (\frac{\partial T}{\partial S})_{Q_c,S_c}=(\frac{\partial^2 T}{\partial S^2})_{Q_c,S_c}=0.
\end{equation}
After some calculation and using \eqref{2.11}, $Q_c$, $S_c$ and the corresponding $T_c$ can be also got as
\begin{eqnarray}
\label{2.16}
  Q_c&=&-\frac{2 \left(-\frac{\lambda }{(\lambda +2) \Lambda }\right)^{\lambda /2}}{\gamma  (\lambda +2) \left(\lambda ^2-1\right)},\\
  \label{2.17}
  S_c&=&-\frac{\pi  \lambda }{(\lambda +2) \Lambda },\\
  T_c&=&\frac{\lambda }{2 \pi  (\lambda +1) \sqrt{-\frac{\lambda }{(\lambda +2) \Lambda }}}.
\end{eqnarray}
Obviously, these critical parameters depend only on the internal parameters of the system $\lambda$ and $\Lambda$. One can also see that $T_c$, $S_c$ increase with $\lambda$ and $Q_c$ decrease with $\lambda$ as shown in Figure (\ref{fig:L7}), $S_c$, $Q_c$ increase with $\Lambda$ and $T_c$ decrease with $\lambda$ as shown in Figure (\ref{fig:L8}). So the results show that the parameter $\lambda$ promote the thermodynamic system to reach the stable state.

For the FPT, we will also check whether Maxwell's equal area law holds
in this thermodynamic system. As is known to all, the first order transition temperature $T^*$ plays an crucial role for the Maxwell's equal area law. Thus in order to get $T^*$, we first plot the curve about the free energy with respect to the temperature $T$, where the free energy is defined by $F = M -TS$. The relation between $F$ with $T$ is plotted in Figure (\ref{fig:L5}). One can see that there is a swallowtail structure, which corresponds to the unstable phase in Figure (\ref{fig:L6}). The non-smoothness of the junction implies that the phase transition is a FPT.  The critical temperature $T^*$ is apparently given by the horizontal coordinate of the junction. From Figure (\ref{fig:L5}), we get $T^*=0.2750$. Substituting this temperature $T^*$ into \eqref{2.11}, one can obtain three values of the entropy, $S_1=0.1563$, $S_2=0.5277$ and $S_3=1.394$. With these values, we can now check the Maxwell's equal area law
\begin{equation}
\label{2.19}
  T^*(S_3-S_1)=\int_{ S_1}^{S_3}T(S,Q)dS.
\end{equation}
After some calculation, we find that both the left and right of \eqref{2.19} equal to $0.3404$ exactly. Thus, our results show that the Maxwell's equal area law is satisfied in this background.

For the   SPT, we will study the critical exponent associated
with the heat capacity definition in \eqref{2.14}. Near the critical point, expanding the temperature as the very small amount $S-S_c$, we find
\begin{equation}
  T=T_c+(\frac{\partial T}{\partial S})_{Q_c,S_c}(S-S_c)+(\frac{\partial^2 T}{\partial S^2})_{Q_c,S_c}(S-S_c)^2+(\frac{\partial^3 T}{\partial S^3})_{Q_c,S_c}(S-S_c)^3+o(S-S_c)^4.
\end{equation}
Using \eqref{2.15}, the second and third terms vanish. Then using \eqref{2.11} ,\eqref{2.16} and \eqref{2.17}, we get
\begin{equation}
  T-T_c=\frac{\lambda }{16 \pi ^4 \left(-\frac{\lambda }{(\lambda +2) \Lambda }\right)^{7/2}}(S-S_c)^3.
  \end{equation}
With the definition of the heat capacity \eqref{2.14}, we further get  $C_Q\sim(T-T_c)^{-2/3}$. So one can find the critical exponent of the heat capacity is $-2/3$, which is the same as the one from the mean field theory in Van der Waals gas-fluid system.

\section{Van der Waals-like Phase transition and critical phenomena of Holographic holographic entanglement entropy}
In this section, our target is to explore whether the HEE have the similar  VDP phase structure and critical phenomena as that of the BHE in massive gravity.  For simplicity, here we only consider the case  $\gamma=-1$. Now we will investigate whether there  is  VDP phase transition in the   HEE - temperature phase plane.

Firstly, we review some basic knowledge about holographic  HEE. Detailed introduction of HEE can refer to Refs.\cite{Ryu-Takayanagi200601,Ryu-Takayanagi200602}.
For a given quantum field theory described by a density matrix $\rho$,  HEE
for a region $A$ and its complement $B$ is
\begin{equation}
  S_A=-Tr_A(\rho_A\ln \rho_A),
\end{equation}
where $\rho_A$ is the reduced density matrix. However,it is usually not easy to get this
quantity in quantum field theory. Fortunately according to AdS/CFT correspondence, Refs.\cite{Ryu-Takayanagi200601,Ryu-Takayanagi200602} propose a very simple
geometric formula for calculating $S_A$ for static states with the area of a bulk minimal
surface as $\partial A$, that is
\begin{equation}
\label{3.2}
  S_A=\frac{Area(\gamma_A)}{4},
\end{equation}
where $\gamma_A$ is the codimension-2 minimal surface according to boundary condition $\partial \gamma_A=\partial A$.

Subsequently using the definition \eqref{3.2}, we will calculate the HEE and study the corresponding phase transition. It is noted that the space on the boundary is a spherical in the AdS black hole in massive gravity and the volume of the space is finite. Thus in order to avoid the  HEE to be affected by the surface that wraps the event
horizon, we will choose a small region as $A$. More precisely, as done in \cite{Zeng201601,Zeng201602,Zeng201603,Zeng201701}, we choose
the region $A$ to be a spherical cap on the boundary given by $\varphi\leq\varphi_0$. Here the area can be written as
\begin{equation}
\label{3.3}
  A=2\pi \int_0^{\varphi_0} \Theta(r(\varphi),\varphi) d\varphi, \Theta= r \sin\varphi\sqrt{\frac{(r')^2}{f(r)}+r^2},
\end{equation}
where $r'=\frac{d r}{d\varphi}$. Then according to the Euler-Lagrange equation, one can get the equation of
motion of $r(\varphi)$, that is
\begin{eqnarray}
  &&0=r'(\varphi)^2[\sin\varphi r(\varphi)^2 f'(r)-2\cos\varphi r'(\varphi)]\\\nonumber
  &&-2r(\varphi)f(r)[r(\varphi)(\sin\varphi r''(\varphi)+\cos\varphi r'(\varphi)-3\sin\varphi r'(\varphi)^2]+4\sin\varphi r(\varphi)^3 f(r)^2.
\end{eqnarray}
After using the boundary conditions $r'(0)=0,r(0)=r_0$, we can get the numeric result
of $r(\varphi)$.

\begin{figure}[tbp]
\centering
\includegraphics[width=.3\textwidth]{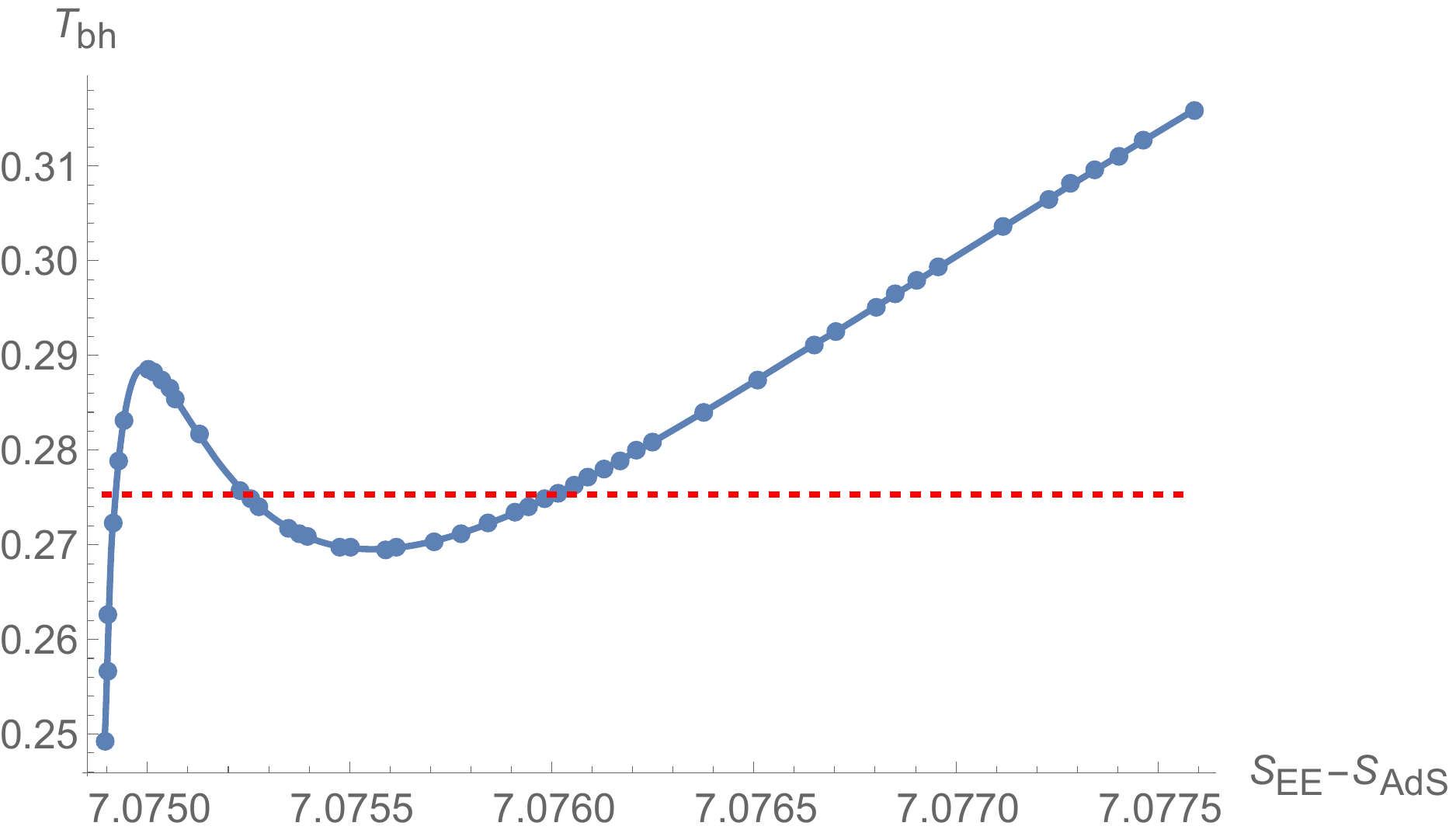}
\hfill
\includegraphics[width=.3\textwidth]{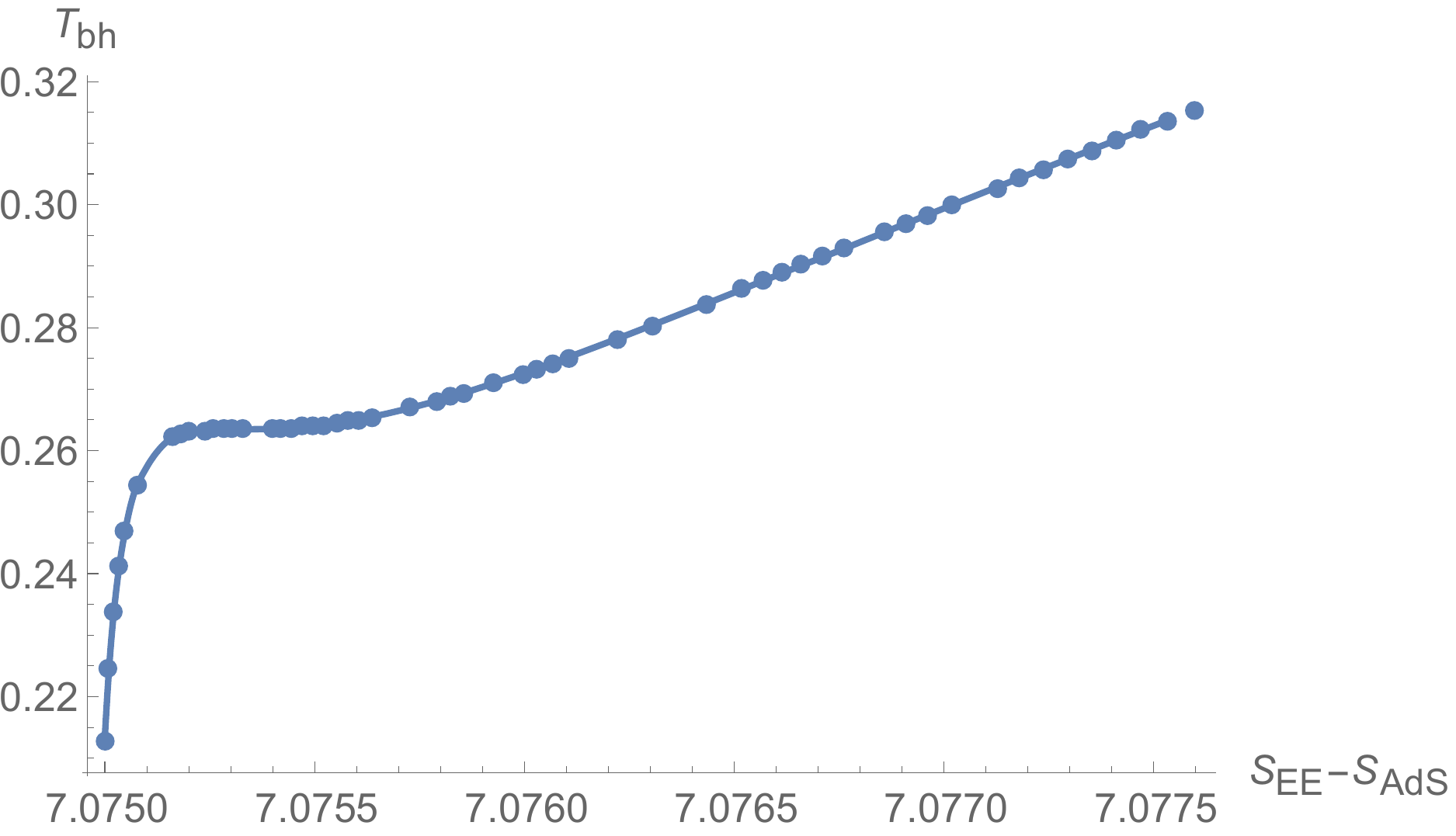}
\hfill
\includegraphics[width=.3\textwidth]{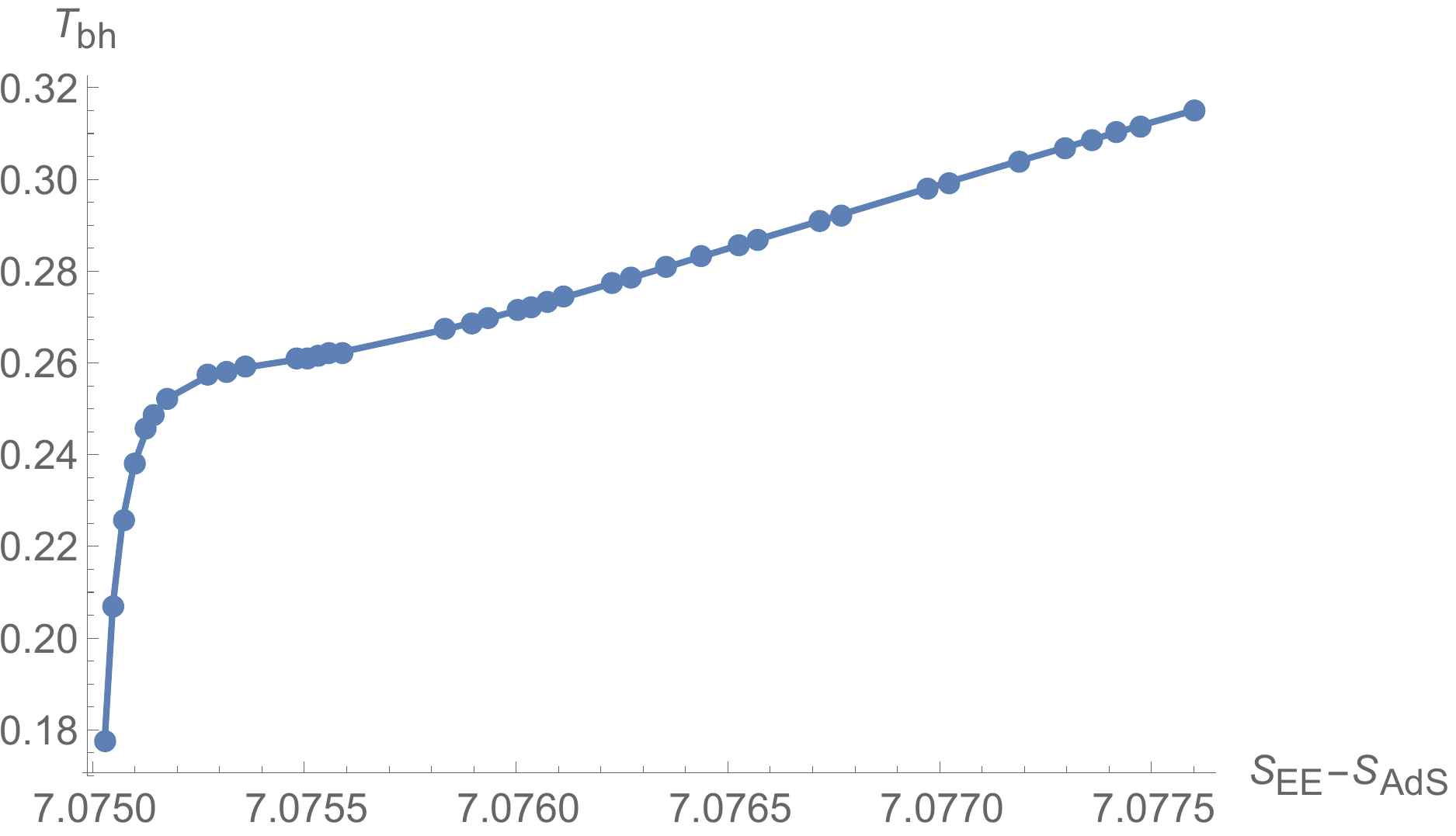}
 \caption{The figure shows T vs $\delta S$ for $\gamma =-1$. Here $\lambda =2.4$, $Q=0.6 Q_c,Q_c,1.2_c$ from left to right}
\label{fig:L9}
\end{figure}

It is worth noting that the HEE should be regularized by subtracting off the  HEE in pure AdS, because of the values of the HEE in the equation \eqref{3.2} is divergent at the boundary. Now let's label the regularized HEE as $\delta S$. Here we choose the size of the boundary region to be $\varphi_0=0.16$  and  set the UV cutoff in the dual field theory to be  $r(0.159)$. The numeric results for $\lambda=2.4$ are shown in Figure(\ref{fig:L9}). One can see that for a given scalar charge $Q$, the relation between the HEE and temperature is similar to that between the black hole   BHE and temperature.  That is to say , the AdS black holes thermodynamic system with the HEE undergos the FPT and   SPT one after another  as the scalar charge $Q$ increase step by step.

For the FPT of the  HEE, we now check whether the
Maxwell's equal area law satisfies. Firstly, for the case $\lambda=2.4,\varphi_0=0.16$, we get an interpolating function of the temperature $T(\delta S)$ using the data obtained numerically. At the first
order phase transition point, we get the smallest and largest roots for the equation $T(\delta S)=0$, that are $\delta S_1 = 7.07492, \delta S_1 = 7.07599$. Then using these values and the equal area law
\begin{equation}
 T^*(\delta S_3-\delta S_1)=\int_{\delta S_1}^{\delta S_3}T(\delta S)dS,
\end{equation}
we find the left equals $0.00029486$ and the right equals $0.00029476$. Obviously both the left and
the right are equal within our numerical accuracy, where the relative error is less than  $0.034 \%$

\begin{figure}[tbp]
 \centering
\includegraphics[width=0.5\textwidth]{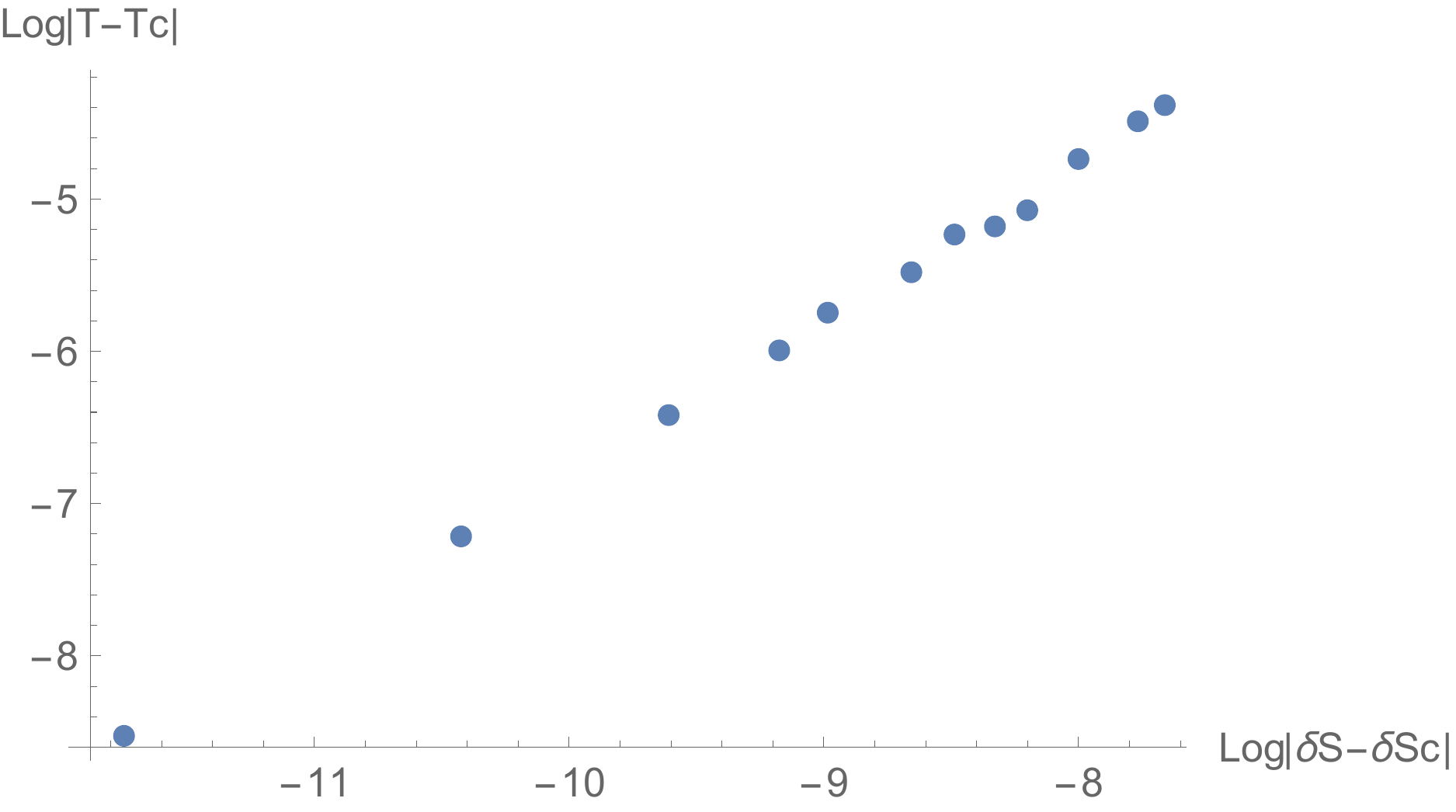}
 \caption{The figure shows $\log|T-T_c|$ vs $\log|\delta S-\delta S_c|$ for $\gamma =-1$. Here $\lambda =2.4$,$Q=Q_c$.}
\label{fig:L10}
\end{figure}

Now let's consider the critical exponent of the   SPT in the holographic  HEE-temperature phase plane. Here comparing with the definition of specific heat capacity $C_Q$ in \eqref{2.14}, one can also define an specific heat capacity for the   HEE as
\begin{equation}
\label{3.6}
  C'_Q=T(\frac{\partial \delta S}{\partial T})_Q.
\end{equation}
Then providing a similar relation of the critical points that in \eqref{2.15} is also working and using \eqref{3.6}, we can get the critical exponent of   SPT of in the HEE- temperature phase. Here we employ the logarithm of the quantities $T-T_c$, $\delta S-\delta S_c$. For $\varphi_0=0.16$, the relation
between $\log |T-T_c|$ and $\log |\delta S-\delta S_c|$ are plotted in Figure (\ref{fig:L10}). The analytical result of these straight lines can be fitted as
\begin{eqnarray}
  \log|T-T_c|=9.023+3.012\log |\delta S-\delta S_c|.
\end{eqnarray}
The result show that the slope is around $3$ with less than relative error $4 \%$, which is consistent with that of the BHE. Then one can find the critical exponent of the specific heat capacity $C'_Q$ is also approximately $-2/3$. That is to say,the HEE has the same   SPT behavior as that of the BHE. Both of them are consistent with the result in the mean field theory of Van der Waals gas-fluid system.


\section{Conclusions}
In this paper, we have investigated the  VDP phase transition with the use of HEE as a probe. Firstly, we investigated the phase diagrams of the BHE in the $T-S$ phase plane and find that the phase structure depends on the scalar charge $Q$ and the parameter $\gamma$ of the AdS black holes in this massive gravity. For the case that $\gamma=1$ or $Q=0$, we find there always exists the Hawking-Page like  phase transition in this thermodynamic system. While for the case  $\gamma=-1$, we found for the small scalar charge $Q$, there is always an unstable black hole thermodynamic system interpolating between the small stable black hole system and large stable black hole system. The thermodynamic transition  for the small hole to the large hole is first order transition and the Maxwell's equal area law is valid. As the scalar charge $Q$ increases to the critical value $Q_c$ in this space time, the unstable black hole merges into an inflection point. We found there is a   SPT at the critical point. When the scalar charge is larger than the critical charge, the black hole is stable always. That is to say, we found that there exists the   VDP gas-fluid phase transition in the $T-S$ phase plane of the AdS black hole in massive gravity.

 The more interesting thing is that we found  the HEE also exhibits the  VDP phase structure in the $T-\delta S$ plane when  $\gamma=-1$ and the scalar charge $Q\neq 0$. In order to confirm this results, we further showed that the Maxwell's equal area law
is satisfied and the critical exponent of the specific heat capacity is consistent with that of the mean field theory of  the VDP gas-fluid system for the HEE system. These results show that the phase structure of HEE is similar to that of  BHE and the  HEE is really a good probe to the phase transition of AdS black holes in Lorentz breaking massive gravity. This also implies that HEE and  BHE exhibit some potential underlying relationship.

\section*{Acknowledgements}

We are grateful to  De-Fu Hou and Hong-Bao Zhang  for their instructive discussions. This work is supported by the National Natural Science Foundation of China (Grant No.11365008, Grant No.11465007, Grant No.61364030).


\end{document}